\documentclass{ws-procs961x669}
\usepackage{slashed}

\newcommand{\sD}{\slashed{D}}
\newcommand{\dd} {\mathrm{d}}
\newcommand{\g}{\sqrt{ g }}
\newcommand{\dc}{\mathcal{D}}
\DeclareMathOperator{\pa}{\partial}
\usepackage{tensor}

\newcommand{\al}{\alpha}
\newcommand{\be}{\beta}

\begin{document}
\title{Model-independent results on 
parity violation in the trace anomaly}
\author{Rémy Larue\footnote{Speaker. E-mail: remy.larue@lpsc.in2p3.fr} }
\address{Laboratoire de Physique Subatomique et Cosmologie,\\
Universit\'{e} Grenoble-Alpes, CNRS/IN2P3, Grenoble INP,\\
53, Avenue des Martyrs 38026 Grenoble, France}
\author{Jérémie Quevillon}
\address{Laboratoire d’Annecy-le-Vieux de Physique Théorique, CNRS — USMB,\\
BP 110 Annecy-le-Vieux, 74941 Annecy, France}
\author{Roman Zwicky}
\address{Higgs Centre for Theoretical Physics, School of Physics and Astronomy,\\
The University of Edinburgh,\\
Peter Guthrie Tait Road, Edinburgh EH9 3FD, Scotland, U.K}

\begin{abstract}
\noindent
Anomalous parity violation in four dimensions would  be significant  for phenomenology  (baryogenesis, gravitational waves) and mathematical physics. Over the past decade, there has been a controversy in the literature as to whether free Weyl fermions give rise to (anomalous) parity violation in the trace of the energy momentum tensor; 
expressed by  the Pontryagin densities $R\tilde R$  and $F\tilde F$  in  the gravity and the gauge sector respectively. In Ref.~\cite{Larue:2023tmu}, we have shown, using path integral methods, that the trace anomaly of a free Weyl fermion does not violate parity (i.e the absence of the Pontryagin density).  In a subsequent work~\cite{Larue:2023qxw} we came to the stronger  conclusion that for any  theory compatible with dimensional regularisation,  the  Pontryagin-terms are equally absent. 
It is the \textit{finiteness} of the diffeomorphism, the Lorentz and the gauge anomalies that prevents  anomalous parity violation.

\end{abstract}


\bodymatter

\section{Introduction}
\label{sec:intro}

It was Capper and Duff\cite{Capper:1974ic} who discovered the Weyl (or trace) anomaly in the 70's. 
 Phenomenological and formal applications were understood over the decades to follow  (see e.g Refs.~\cite{Duff:1993wm,Duff:2020dqb}). In analogy to  the axial anomaly that manifests itself in the divergence of the  axial current, the Weyl anomaly appears in the divergence of the dilatation current, which incidentally is the trace of the energy-momentum tensor (EMT)
\begin{equation}
\mathcal{A}_{\text{Weyl}}=g_{\mu\nu}\langle T^{\mu\nu}\rangle\;.
\end{equation}
In the remainder of this section  we review 
scale invariance in flat and curved spacetime,  their respective anomalies  and how parity violation may occur. In Section~\ref{sec:Weylfermion} we investigate the presence of parity violation in a Weyl fermion theory. 
Then, in Section~\ref{sec:modelindep} we broaden this investigation to any theory that is compatible with dimensional regularisation, which includes spin 3/2 and spin 2\,\cite{Christensen:1978gi,Christensen:1978md,Perry:1978jj,Critchley:1978ki,Critchley:1978kb,Yoneya:1978gq,Birrell:1982ix}. Our main conclusion is that the \textit{finiteness} of the diffeomorphism and Lorentz anomalies is sufficient to rule out the presence of parity violation in the Weyl anomaly. Finally, we summarise our results in Section~\ref{sec:conclusion}.

\paragraph{Flat spacetime}

In flat spacetime, massless theories bear no intrinsic scale, and as a result exhibit an invariance under the rescaling of spacetime and of the fields. Let us consider for example a vector-like Dirac fermion with action
\begin{equation}\label{eq:Sfermion}
S=\int\dd^dx\,\bar\psi i\sD\psi\;,
\end{equation}
and covariant derivative
\begin{equation}
D_\mu\psi=(\pa_\mu+iV_\mu)\psi\;,
\end{equation}
$V$ being some vector gauge field. Under a scale transformation with constant parameter $\sigma$,  spacetime variables and  fields transform according to (see e.g Ref.~\cite{Fujikawa:2004cx})
\begin{alignat}{4}\label{eq:scaletransfo}
&x'_\mu&\;=\;&e^{\sigma}x_\mu\;,&&\frac{\pa}{\pa x'_\mu} &\;=\;&e^{\sigma}\frac{\pa}{\pa x_\mu}\;,\nonumber\\
&\psi'(x')&\;=\;&e^{-\frac{d-1}{2}\sigma}\psi(x)\;,\quad &&\bar\psi'(x')&\;=\;&e^{-\frac{d-1}{2}\sigma}\bar\psi(x)\;,\quad V'_\mu(x')=e^{-\sigma}V_{\mu}(x)\;,
\end{alignat}
and one can readily verify that the action \eqref{eq:Sfermion} is indeed scale invariant, with conserved dilation current $J_D^\mu$ 
\begin{equation}
\partial \cdot J_D = \tensor{T}{^\mu_\mu}=0\;.
\end{equation}
The presence of a mass term $m\bar\psi\psi$ in the action would naturally break the scale invariance, due to the introduction of an explicit scale.

\paragraph{Curved spacetime}
Our goal is to analyse the scale invariance of a theory coupled to gravity. Fermions couple to gravity via the vierbein fields $\tensor{e}{^a_\mu}$. The latter translate between the manifolds metric $g$ and 
the  Minkowski metric $\eta$ 
\begin{equation}
g_{\mu\nu}=\tensor{e}{^a_\mu}\tensor{e}{^b_\nu}\eta_{ab}\;,
\end{equation}
 and are used to define the spin-connection $\tensor{\omega}{_\mu^a_b}=-\tensor{e}{_b^\nu}\pa_\mu\tensor{e}{^a_\nu}+\tensor{e}{^a_\nu}\tensor{e}{^\rho_b}\Gamma^\nu_{\mu\rho}$ (where $\Gamma$ are the Christoffel symbols).
The action is then given by
\begin{equation}\label{eq:Sfermiongrav}
S=\int\dd^dx\,\g\,\bar\psi i\tensor{e}{^\mu_a}\gamma^a D_\mu\psi\;,\quad\quad D_\mu\psi=(\pa_\mu+\omega_\mu+iV_\mu)\psi\;,
\end{equation}
which is invariant under the symmetries of General Relativity, namely the diffeomorphisms  (or general coordinate transformations) and the Lorentz transformations (given in  Appendix \ref{app:transfo}). When the fermion is solution to the classical equations of motion $\delta S/\delta\psi=0$ and $\delta S/\delta\bar\psi=0$, we obtain the conserved currents for each of these symmetries. If we denote by $\delta^d_{\xi}$ ($\delta^L_\alpha$) the infinitesimal diffeomorphism (resp. Lorentz) transformation of parameter $\xi_\mu(x)$ (resp. $\alpha_{ab}(x)=-\alpha_{ba}(x)$), we have
\begin{align}
&\delta^d_\xi S=\int\dd^dx\,\xi^\nu\left(D^\mu T_{\mu\nu}-\omega_{\nu ab}T^{ab}\right)=0\nonumber\\
&\delta^L_\alpha S=-\int\dd^dx\alpha_{ab}T^{ab}=0\;,
\end{align}
where the EMT is defined by
\begin{equation}
\tensor{T}{^\mu_a}=\frac{1}{\g}\frac{\delta S}{\delta \tensor{e}{^a_\mu}}\;,\quad\quad T^{\mu\nu}=g^{\nu\rho}\tensor{e}{^a_\rho}\tensor{T}{^\mu_a}\;.
\end{equation}
Let us return to the task at hand, that is evaluate the scale invariance of the action \eqref{eq:Sfermiongrav}. One may be tempted to perform the same transformation as the one in flat spacetime \eqref{eq:scaletransfo}, which involves both a transformation of the fields and of the spacetime variable $x$. In General Relativity, a transformation of the spacetime variable is a diffeomorphism, hence the scale transformation \eqref{eq:scaletransfo} in curved spacetime is a mix of a rescaling of the fields and of a diffeomorphism, and is not well-suited \cite{Fujikawa:2004cx}. The adequate  operation  is obtained by trading the transformation of the spacetime variable for a transformation of  the metric, or equivalently the vierbein: this is called a Weyl transformation. Denoting by the $\delta^W_\sigma$ the infinitesimal Weyl transformation of parameter $\sigma(x)$ in $d$ dimensions, we have
\begin{alignat}{6}
&\delta^{W}_\sigma\tensor{e}{^\mu_a} &\;=\;& -\sigma\, \tensor{e}{^\mu_a}\;, \quad   & & \delta^{W}_\sigma e&\;=\;& d\,\sigma\, e
\;, \quad   & & \delta^{W}_\sigma \omega_\mu&\;=\;& \frac{d-1}{2}(\pa_\mu\sigma) \;,
 \nonumber \\[0.1cm]
&\delta^{W}_\sigma\psi &\;=\;& -\frac{d-1}{2}\sigma\,\psi \;, \quad   & & \delta^{W}_\sigma\bar\psi &\;=\;&
-\frac{d-1}{2}\sigma\,\bar\psi \;,\label{c3eq:Weyltransfo}
\end{alignat}
where $e=\g$ is the vierbein determinant.
The massless fermionic action is invariant under this transformation, and when the fermion is solution to its equations of motion we find
\begin{equation}
\delta^W_\sigma S=\int\dd^dx\,\g\,\tensor{T}{^\mu_\mu}=0\;.
\end{equation}

\paragraph{Scale vs conformal vs Weyl transformation}
Let us clarify a point that is often source of confusion in the literature. As emphasised above, the global scale transformation represents a rescaling of the theory in flat spacetime. In curved spacetime the concept of global scale transformation 
is not adequate. It is the  Weyl transformation that generalises  the scale transformation (to any spacetime manifold).

Closely related to those is the conformal transformation, which is a diffeomorphism along a conformal Killing vector field $\xi_\mu(x)$, such that the metric is effectively rescaled 
\begin{equation}
\delta^d_\xi g_{\mu\nu}=\Omega(x)g_{\mu\nu}\;,\quad \Omega(x)=\frac{2}{d}(D_\mu\xi^\mu)\;.
\end{equation}
This cannot be applied in any spacetime since only manifolds of constant curvature admit conformal Killing vector fields. For example in flat spacetime, the conformal group decomposes into: translations, rotations, the special conformal transformation, and the scale transformation mentioned above.

\paragraph{Weyl anomaly}
Similarly as for the axial symmetry, the Weyl symmetry is broken at the quantum level. The reason behind this is that the quantisation procedure introduces an intrinsic scale, which is the renormalisation scale.

Let us consider the quantum effective action in Euclidean time
\begin{equation}
W(d)=-\log \int \dc \phi\,e^{-S}\;.
\end{equation}
It is regularised using dimensional regularisation: $d=4-2\epsilon$. The renormalised, hence finite, theory is obtained by introducing the counterterms which are of the form
\begin{equation}\label{eq:Wctform}
W_{ct}=\frac{1}{d-4}\int\dd^dx\,\g\,\mathcal{P}\;,
\end{equation}
where $\mathcal{P}$ is a local polynomial in the background fields, and the renormalised effective action reads
\begin{equation}
W_{ren}(d)=W(d)+W_{ct}(d)\;.
\end{equation}
Even though the theory is classically Weyl invariant, i.e $\delta^W_\sigma S=0$, the quantum theory is in general not invariant and we have\cite{Larue:2023qxw}
\begin{equation}
\delta^W_\sigma\lim_{d\to4} W_{ren}(d)=\lim_{d\to4}\delta^W_\sigma W_{ren}(d)=\int\dd^4x\,\g\,\mathcal{A}_{\text{Weyl}}\;,
\end{equation}
where the $d\to4$ limit and the Weyl variation commute since $W_{ren}(d\to4)$ is finite by construction. 
The quantity  $\mathcal{A}_{\text{Weyl}}$ is finite and is called the Weyl anomaly, discovered by Capper and Duff in the 70s\cite{Capper:1974ic}. For a classically Weyl invariant theory, and using dimensional regularisation, it can be expressed as\footnote{Quantities inside Vacuum Expectation Values (VEV) live in $d=4-2\epsilon$ dimensions whereas quantities outsides the VEV live in 4 dimensions. The superscript $(d)$ indicates the dimensions of  the metric, which is $d=4-2\epsilon$ when not specified.}\cite{DESER197645,DUFF1977334,Duff:1993wm,Duff:2020dqb}
\begin{equation}\label{eq:AWeylCFT}
\mathcal{A}_{\text{Weyl}}=g_{\mu\nu}\langle T^{\mu\nu}\rangle=\frac{1}{\g}\left[g^{(d)}_{\mu\nu}\frac{W_{ct}(d)}{\delta g^{(d)}_{\mu\nu}}\right]_{d=4}\;.
\end{equation}
The anomaly is also present in a theory with explicit Weyl breaking, i.e $\delta^W_\sigma S\neq0$, as one would expect, and is often defined as~\cite{Duff:1993wm,Duff:2020dqb,Casarin:2021fgd}
\begin{equation}
\mathcal{A}_{\text{Weyl}}=g_{\mu\nu}\langle T^{\mu\nu}\rangle-\langle \tensor{T}{^\mu_\mu} \rangle\;,
\end{equation}
which was formally shown to hold in dimensional regularisation and at one-loop\footnote{Beyond one-loop considerations can be found in Ref.~\cite{Larue:2023qxw,Bertolini:2024vwu}. The expression of the anomaly in different regularisation schemes was in recently investigated in Ref.~\cite{Ferrero:2023unz}.} in Ref.~\cite{Larue:2023qxw}, as well as its equivalent formulation\footnote{\label{footnote}This expression involving the counterterms holds in minimal subtraction and has to be slightly modified otherwise. Nonetheless the analysis of Ref.~\cite{Larue:2023qxw} remains valid (see Appendix A therein for more details).}
\begin{equation}\label{eq:AWeylNoCFT}
\mathcal{A}_{\text{Weyl}}=\frac{1}{\g}\left[g^{(d)}_{\mu\nu}\frac{\delta W_{ct}(d)}{\delta g^{(d)}_{\mu\nu}}\right]_{\text{fin}}\;,
\end{equation}
where the subscript finite indicates that only the finite piece contributes. Note that in the presence of explicit breaking the anomaly is not integrable, that is to say it does not respect the Wess-Zumino consistency conditions \cite{Wess:1971yu,Larue:2023qxw}.

As an example, the Weyl anomaly of the massless vector-like Dirac fermion \eqref{eq:Sfermiongrav} is\cite{Bertlmann:1996xk,Fujikawa:2004cx}
\begin{equation}\label{eq:Avectorlike}
\mathcal{A}_{\mathrm{Weyl}}^{\mathrm{Dirac}} = \frac{1}{16\pi^2}\bigg(\frac{1}{72}R^2-\frac{1}{45}R_{\mu\nu}R^{\mu\nu}-\frac{7}{360}R_{\mu\nu\rho\sigma}R^{\mu\nu\rho\sigma} 
-\frac{1}{30}\Box R +\frac{2}{3}F^2
\bigg)\;,
\end{equation}
where  $F_{\mu\nu}$ the gauge field strength.

\paragraph{Weyl anomaly and parity violation}

A general analysis reveals that we may reduce the form of the Weyl anomaly in 4 dimensions to the following
\begin{equation}\label{eq:AWeylgeneric}
\mathcal{A}_{\mathrm{Weyl}}=a\,E+b\,R^2+c\,W^2+d\,\Box R+e\,R\tilde R+f\,F^2+g\,F\tilde F\;,
\end{equation}
where the Euler density is $E=R_{\mu\nu\rho\sigma}R^{\mu\nu\rho\sigma}-4R_{\mu\nu}R^{\mu\nu}+R^2$, the Weyl tensor squared is $W^2=R_{\mu\nu\rho\sigma}R^{\mu\nu\rho\sigma}-2R_{\mu\nu}R^{\mu\nu}+\frac{1}{3}R^2$ and the parity violating topological Pontryagin densities are\footnote{Note that they also respect the Wess-Zumino consistency conditions \cite{Bonora:1985cq}.}
\begin{equation}
R\tilde R=\frac{1}{2}\epsilon^{\mu\nu\rho\sigma}\tensor{R}{^\alpha^\beta_\mu_\nu}\tensor{R}{_\alpha_\beta_\rho_\sigma}\;,\quad\quad F\tilde F=\frac{1}{2}\epsilon^{\mu\nu\rho\sigma}F_{\mu\nu}F_{\rho\sigma}\;.
\end{equation}
These $P$--odd operators are  phenomenologically relevant (see for example the discussion in Ref.~\cite{Nakayama:2012gu}, or baryogenesis and gravitational waves as reviewed in Ref.~\cite{Alexander:2009tp}), and could be observed experimentally \cite{Chu:2022bhj}.

Moving forward, our goal will be to analyse the presence of these parity-violating terms in the trace anomaly. 

\section{Weyl anomaly of Weyl fermions}
\label{sec:Weylfermion}

Although vector-like fermions generate a parity-even trace anomaly \eqref{eq:Avectorlike}, parity-violation may still arise in other theories. In particular, the case of a left-handed fermion with action
\begin{equation}\label{eq:lefthandedDirac}
S_L=\int\dd^4x\,\bar\psi\,i\sD P_L\psi\;, \quad P_L=\frac{1-\gamma_5}{2} \;,
\end{equation}
has been the subject of debates over the past decade. The $R\tilde R$-term was found to be non-vanishing in Ref.~\cite{Bonora:2014qla}. In the following years, some authors have found a non-vanishing coefficient $e$  (defined in Eq.~\eqref{eq:AWeylgeneric}) \cite{Bonora:2014qla,Bonora:2015nqa,Bonora:2017gzz,Bonora:2018ajz,Bonora:2018obr,Bonora:2022izj,Liu:2022jxz,Liu:2023yhh}  whereas others have found it to be vanishing  \cite{Bastianelli:2016nuf,Bastianelli:2019zrq,Frob:2019dgf,Abdallah:2021eii,Abdallah:2023cdw}.  Weyl fermions are subtle and probe spacetime in their own way; 
in $d = 2\;(\textrm{mod } 4)$ they give rise to gravitational (diffeomorphism and Lorentz) anomalies \cite{Alvarez-Gaume:1983ihn}, or more accurately  to Lorentz anomalies only \cite{LEUTWYLER198565,Leutwyler:1985ar}.
In the specific determinations of $e \neq 0$ there is something similarly unsettling in that the authors found  it to be purely imaginary in Lorentzian signature. This in fact implies that its contribution is $CPT$-violating since $T \circ i = - i$ which would indicate a $CPT$ anomaly. Whereas it was noted that an imaginary $e$ would violate unitarity  \cite{Bonora:2014qla}, the $CPT$-violation itself seems to have been overlooked.  This would either mean that such theories have to be discarded  \cite{Bonora:2014qla} or supplemented by new particles such as three right-handed neutrinos in the Standard Model.

The determination of the Weyl anomaly of Weyl fermions is non-trivial for several reasons. The main difficulty that often arises when computing anomalies is that several currents are entangled by the regularisation. For the case at hand, the symmetries of gravity, Lorentz and diffeomorphism invariance, are entangled with the Weyl symmetry since it transforms the metric. Although, diffeomorphism and Lorentz anomalies vanish in $d=4$\;\cite{Alvarez-Gaume:1983ihn}, the computation should stay free from spurious anomalies, i.e anomalies that can be removed by local polynomial counterterms. A wrong choice of regularisation may lead to such spurious diffeomorphism and/or Lorentz anomalies, and yield an incorrect Weyl anomaly. This is similar to the distribution of the anomaly between vector and axial symmetries in a vector-axial gauge theory (see e.g Ref.\cite{Filoche:2022dxl}).

Second, the left-handed Dirac fermion from \eqref{eq:lefthandedDirac} has an ill-defined propagator since $i\sD P_L$ is not invertible. Several authors chose to paliate this problem by introducing a right-handed spectator fermion. Although this poses no problem in a gauge theory and has been used extensively \cite{Bardeen:1969md,Alvarez-Gaume:1984zlq}, it falls short  with gravity since the right-handed spectator is not coupled to gravity and therefore  violates Lorentz invariance~\cite{Larue:2023tmu}. This problem is circumvented easily by considering instead a two-component Weyl fermion
\begin{equation}
S_L=\int\dd^4x\,\g\,\bar\psi_L\,i\bar\sigma^\mu D_\mu\psi_L\;,\quad\quad D_\mu\psi_L=(\pa_\mu+\omega_L+i V_\mu)\;,
\end{equation}
where $\bar\sigma^\mu = \tensor{e}{^\mu_a}\bar \sigma^a$, $\bar\sigma^a = (1, -\vec{\sigma})$ with $\vec{\sigma}$ the Pauli matrices; $\omega_L$ is the projected spin-connection such that $\omega_{\mu ab} P_L=\mathrm{diag}(0,\omega_{L,\mu ab})$. This provides a well-defined propagator: $(i\bar\sigma^\mu D_\mu)^{-1}$.

Third, 
the diagrammatic approach in curved spacetime requires to linearise the metric around a Minkowski background,  resulting in  heavy and hard to track computations.  Since  the computation can only be carried out up to some order in the metric,
conservation of diffeomorphism and Lorentz invariance cannot be fully verified. Besides, the covariantisation (which consists in inferring the result in terms of curvature invariants from the incomplete metric expansion) is endangered by this potential violation of diffeomorphism and Lorentz invariance.

Lastly, although the path integral approach can be performed around any background metric (i.e no need to linearise), it suffers from an ill-defined functional determinant\cite{Alvarez-Gaume:1983ihn} as we now show. We would like to define the path integral as
\begin{equation}\label{eq:WSL}
W=-\log\int\dc\tilde{\bar\psi}_L\dc\tilde{\psi}_L\,e^{-S_L}=-\log\det\,i\bar\sigma^\mu D_\mu\;,
\end{equation}
where we defined the diffeomorphism-invariant measure with $\tilde\psi_L=g^{1/4}\psi_L$ and $\tilde{\bar\psi}_L=g^{1/4}\bar\psi_L$\cite{Fujikawa:1980rc,Toms:1986sh,Fujikawa:2004cx,Larue:2023tmu}.
The functional determinant is then understood as the product of the eigenvalues of $i\bar\sigma^\mu D_\mu$, excluding zero-modes. However, the eigenvalue equation
\begin{equation}\label{eq:eigenvalue}
i\bar\sigma^\mu D_\mu\phi=\lambda\phi
\end{equation}
is meaningless, since $i\bar\sigma^\mu D_\mu$ maps left-handed chiralities onto right-handed ones, and \eqref{eq:eigenvalue} is an equality between fermions of different chiralities. The way out proposed by Leutwyler and Mallik\cite{Leutwyler:1984de,LEUTWYLER198565,Leutwyler:1985ar,Leutwyler:1985em}, is to consider the variation of the functional determinant, which is formally given by
\begin{equation}
\delta W=\delta\log\det i\bar\sigma^\mu D_\mu=\mathrm{Tr}\,\delta(i\bar\sigma^\mu D_\mu)(i\bar\sigma^\mu D_\mu)^{-1}\;,
\end{equation}
which is well-defined since the operator on the right-hand side maps fermions onto fermions of the same chirality. This quantity gives the quantum breaking of a transformation (if the symmetry is classically verified) and, after regularisation, the anomaly. In Ref.~\cite{Larue:2023tmu}, we built upon this work using a proper time regularisation,  to compute the Weyl anomaly, as well as diffeomorphism and Lorentz anomalies, to ensure the absence of spurious anomalies.

As an important remark, since the regularisation may \textit{a priori} break diffeomorphism and Lorentz invariance, conventional computation methods cannot be applied  straightforwardly, since the momentum representation in curved spacetime is often introduced by relying on a specific choice of coordinates\cite{Bunch:1979uk}. We use the method proposed in Ref.~\cite{Larue:2023uyv}, that can be used to compute functional determinants in a coordinate-independent manner, and allows us to explicitly verify the conservation of diffeomorphism and Lorentz invariance.

The result from\cite{Larue:2023tmu,Larue:2023qxw} is that the Weyl anomaly of a Weyl fermion is half that of a Dirac fermion
\eqref{eq:Avectorlike} 
\begin{equation}
\label{eq:AWeyl}
\mathcal{A}^{\mathrm{Weyl}}_{\mathrm{Weyl}}= \frac{1}{32\pi^2}\bigg(\frac{1}{72}R^2-\frac{1}{45}R_{\mu\nu}R^{\mu\nu}-\frac{7}{360}R_{\mu\nu\rho\sigma}R^{\mu\nu\rho\sigma} 
-\frac{1}{30}\Box R +\frac{2}{3}F^2
\bigg)\;,
\end{equation}
while verifying that the effective action is indeed diffeomorphism and Lorentz invariant, i.e free from spurious anomalies. Consequently, the Pontryagin densities $R\tilde R$ and $F\tilde F$ are absent.

\section{Model-independent constraints on the trace anomaly}
\label{sec:modelindep}

We now know that Weyl fermions do not give rise to a parity-violating Weyl anomaly, but the question remains as to whether other models could. In Ref.~\cite{Larue:2023qxw}, we proposed a model-independent analysis of the Weyl anomaly under the light of diffeomorphism and Lorentz anomalies.\footnote{In another approach, parity-odd terms are investigated in correlation functions in Refs.~\cite{Coriano:2023cvf,Coriano:2024wsb} in a conformal field theory. We stress that these are consistency constraints and not a proof of existence.}  It leads to powerful and very generic constraints on its form, as we will see below.

The parity-even part of the Weyl anomaly was already found by Duff\cite{DUFF1977334} to be constrained in the case of a classically Weyl invariant theory. It was found that $b=0$ in \eqref{eq:AWeylgeneric}, which amounts to enforcing the Wess-Zumino consistency conditions\cite{Bonora:1983ff,Bonora:1985cq,Osborn1991WeylCC,Cappelli:1988vw,Cappelli:1990yc}, and that the coefficient of the $\Box R$-term is fixed with respect to the others. The result can be summarised as follows: for a classically Weyl invariant theory, and considering only $P$-even operators, we have
\begin{equation}
\mathcal{A}_{\mathrm{Weyl}}=a_1\, R^2+a_2\,R_{\mu\nu}R^{\mu\nu}- (3\, a_1 +a_2) \,R_{\mu\nu\rho\sigma}R^{\mu\nu\rho\sigma}- (4\, a_1+a_2) \,\Box R\;.
\end{equation}
Our analysis includes theories with 
explicitly broken Weyl synmetry and $P$-odd operators to extend these constraints. It relies on the facts that
\begin{itemize}
\item The Weyl anomaly, with and without explicit breaking, can be fully determined by the counterterms $W_{ct}$, according to \eqref{eq:AWeylCFT} and \eqref{eq:AWeylNoCFT}.
\item The diffeomorphism and Lorentz anomalies are topological, hence finite and unaffected by the counterterms~\cite{Adler:1969er,Alvarez-Gaume:1983ihn}. In other words, the counterterms must be both diffeomorphism and Lorentz invariant: $\delta^d_\xi W_{ct}=\delta^L_\alpha W_{ct}=0$.
\end{itemize}
Now, the counterterms to the EMT are given by
\begin{equation}
\mathcal{T}_{ct}^{\mu\nu}=\frac{1}{\g}\tensor{e}{^\nu_a}\frac{\delta W_{ct}}{\delta \tensor{e}{^a_\mu}}\;,
\end{equation}
and diffeomorphism and Lorentz constraints combined read
\begin{equation}\label{eq:diffeoLorentzconstraints}
D_\nu\mathcal{T}_{ct}^{\mu\nu}=0\; \text{ and }\mathcal{T}_{ct}^{\mu\nu}=\mathcal{T}_{ct}^{\nu\mu}\;,
\end{equation}
and the Weyl anomaly given by \eqref{eq:AWeylCFT} and \eqref{eq:AWeylNoCFT} reads
\begin{alignat}{3}
& \mathcal{A}_{\mathrm{Weyl}} &\;=\;&  [g^{(d)}_{\mu\nu}\mathcal{T}_{ct}^{\mu\nu}]_{d=4}  \quad \quad & &    \text{classically Weyl invariant} \;, \nonumber \\[0.1cm]
& \mathcal{A}_{\mathrm{Weyl}} &\;=\;& [g^{(d)}_{\mu\nu}\mathcal{T}_{ct}^{\mu\nu}]_{\text{fin},d=4}  \quad \quad   & & \text{classically Weyl non-invariant} \;,\label{eq:WeylTct}
\end{alignat}
where $g^{(d)}_{\mu\nu}$ denotes the metric in $d=4-2\epsilon$ dimensions.

\subsection{Generic ansatz}
The next and decisive task is to apply the diffeomorphism and Lorentz constraints on $\mathcal{T}_{ct}$ to obtain constraints on the Weyl anomaly. To that end, we define the most generic ansatz to obtain model-independent results.

The $P$-even part of $\mathcal{T}_{ct}$ is easily dealt with, it suffices to introduce a set of independent operators built with diffeomorphism and Lorentz invariant quantities.\footnote{Note that enforcing $\mathcal{T}_{ct}$ to be diffeomorphism and Lorentz invariant does not imply that the diffeomorphism and Lorentz anomalies vanish, but that they are restricted to their covariant form\cite{Bertlmann:1996xk}.} A possible (i.e. complete) choice is
\begin{align}
\mathcal{T}^{\alpha\beta}_{ct,\text{even}}=&\frac{1}{d-4}\bigg\{a_1\,g^{\alpha\beta}R^2+a_2\,g^{\alpha\beta}R_{\mu\nu}R^{\mu\nu}+a_3\,g^{\alpha\beta}R_{\mu\nu\rho\sigma}R^{\mu\nu\rho\sigma}+a_4\,g^{\alpha\beta}\Box R\nonumber\\
&+b_1\,R R^{\alpha\beta}+b_2\, R^{\alpha\lambda}\tensor{R}{^\beta_\lambda}+b_3\,R_{\mu\nu}\tensor{R}{^\mu^\alpha^\nu^\beta}+b_4\,R^{\alpha\lambda\mu\nu}\tensor{R}{^\beta_\lambda_\mu_\nu}\nonumber\\
&+c_1\, D^{\alpha} D^{\beta}R+c_2\,\Box R^{\alpha\beta}\bigg\} \;,
\end{align}
with  operators defined in $d=4-2\epsilon$ dimensions.

The $P$-odd part is more subtle  to deal with. The $P$-odd part of the counterterms must be of the form
\begin{equation}
W_{ct,\text{odd}} = \frac{1}{d-4}\int\dd^dx \,\g\,\mathcal{L}_{\mathrm{odd}}(d)\;,
\end{equation}
where $\mathcal{L}_{\mathrm{odd}}(d)$ is some local $P$-odd polynomial defined in $d=4-2\epsilon$ dimensions. In particular, we must have $\mathcal{L}_{\mathrm{odd}}(4)  \propto R\tilde R$ as this is the only possible term in $d=4$. Owing to its topological nature one has
\begin{equation}\label{eq:deltaLodd4}
\frac{\delta}{\delta g^{(4)}_{\alpha\beta}}\int\dd^4x\,\sqrt{g^{(4)}}\,\mathcal{L}_{\mathrm{odd}}(4)=0\; .
\end{equation}
One might think that that in $d$ dimensions we should then have 
\begin{equation}
\frac{\delta}{\delta g^{(d)}_{\alpha\beta}}\int\dd^dx\,\sqrt{g^{(d)}}\,\mathcal{L}_{\mathrm{odd}}(d)=\mathcal{O}(d-4)\;,
\end{equation}
however, this would be assuming that extensions of the Pontryagin density to $d=4-2\epsilon$ dimensions are continuous around $d=4$. This statement is not ensured due to its topological nature, for example the topological Euler density is known to be discontinuous around $d=4$\,\cite{Hennigar:2020lsl,Gurses:2020ofy}.
We may however write
\begin{equation}\label{eq:deltaLodd}
\frac{1}{\g}\frac{\delta}{\delta g_{\alpha\beta}}\int\dd^dx\,\g\,\mathcal{L}_{\mathrm{odd}}(d)=(d-4) \mathcal{V}^{\alpha\beta}(4)+\mathcal{U}^{\alpha\beta}(d)+\mathcal{O}\left((d-4)^2\right)\; ,
\end{equation}
where $\mathcal{U}^{\alpha\beta}(d)$ is discontinuous around $d=4$ and $\mathcal{V}^{\alpha\beta}(4)$ is defined in $4$ dimensions. Upon using Bianchi identities and the intrinsically 4-dimensional Schouten identity\cite{Remiddi:2013joa,Chala:2021cgt,Chung:2022ees}, it can be shown that the minimal set of independent operators for $\mathcal{V}^{\alpha\beta}$ reduces to
\begin{equation}
\mathcal{V}^{\alpha\beta}(4)=e\,g^{\alpha\beta}R\tilde R\;.
\end{equation}
Similarly we identify a set of independent operators for $\mathcal{U}^{\alpha\beta}(d)$
(for which   the Schouten identities do not apply). We obtain
\begin{equation}
\mathcal{U}^{\mu\nu}(d)=e_1\, g^{\mu\nu}R\tilde R(d)+e_2\,P^{\mu\nu}(d)+e_3\,Q^{(\mu\nu)}(d)+e_4\,S^{(\mu\nu)}(d)\;,
\end{equation}
where round brackets denote symmetrisation $t^{(\mu\nu)}=\frac{1}{2}(t^{\mu\nu}+t^{\nu\mu})$ and
\begin{alignat}{4}
&g_{\al\be}R\tilde R(d)&\;=\;&g_{\al\be}\frac{1}{2}\epsilon^{\mu\nu\rho\sigma}R_{\gamma\delta\mu\nu}\tensor{R}{^\gamma^\delta_\rho_\sigma}\;,\quad\quad &P_{\alpha\beta}(d)&\;=\;&\epsilon^{\mu\nu\rho\sigma}R_{\alpha\lambda\mu\nu}\tensor{R}{_\beta^\lambda_\rho_\sigma}\nonumber\;, \\[0.1cm]
&Q_{\alpha\beta}(d)&\;=\;&\tensor{\epsilon}{_\alpha^\nu^\rho^\sigma}R_{\beta\nu\gamma\delta}\tensor{R}{_\rho_\sigma^\gamma^\delta}\;,\quad\quad &S_{\alpha\beta}(d)&\;=\;&\tensor{\epsilon}{_\alpha^\nu^\rho^\sigma}R_{\beta\lambda\rho\sigma}\tensor{R}{_\nu^\lambda}\nonumber\, .
\end{alignat}
Let us remark that the coefficients $e_i$ are not independent since they are constrained by the topological nature of $\mathcal{L}_{\text{odd}}(4)$ Eq.~\eqref{eq:deltaLodd4}, which implies that $\mathcal{U}^{\alpha\beta}(4)=0$ and thus
\begin{equation}\label{eq:Ualbe(d=4)}
2\,e_1+e_2+e_3=0\;,
\end{equation}
upon using the Schouten identity: $P_{\mu\nu}(4)=Q_{\mu\nu}(4)=\frac{1}{2}R\tilde R(4)$ and $S_{\mu\nu}(4)=0$. Note that we have indeed constructed a tensor that is discontinuous around $d=4$ since $\mathcal{U}^{\alpha\beta}(4)=0$, but $\mathcal{U}^{\alpha\beta}(d=4-2\epsilon)\neq\mathcal{O}(\epsilon)$.

Finally, our ansatz reads
\begin{align}
\label{eq:Talphabeta}
\mathcal{T}_{ct}^{\al\be}=\frac{1}{\g}\frac{\delta W_{ct}}{\delta g_{\al\be}}=&e\,g^{\al\be}R\tilde R+\frac{1}{d-4}\bigg\{e_1\,g^{\al\be}R\tilde R+e_2 \,P^{\al\be}+e_3\, Q^{(\al\be)}+e_4 \,S^{(\al\be)}\nonumber\\
& +a_1\,g^{\alpha\beta}R^2+a_2\,g^{\alpha\beta}R_{\mu\nu}R^{\mu\nu}+a_3\,g^{\alpha\beta}R_{\mu\nu\rho\sigma}R^{\mu\nu\rho\sigma}+a_4\,g^{\alpha\beta}\Box R\nonumber\\
&+b_1\,R R^{\alpha\beta}+b_2\, R^{\alpha\lambda}\tensor{R}{^\beta_\lambda}+b_3\,R_{\mu\nu}\tensor{R}{^\mu^\alpha^\nu^\beta}+b_4\,R^{\alpha\lambda\mu\nu}\tensor{R}{^\beta_\lambda_\mu_\nu}\nonumber\\
&+c_1\, D^{\alpha} D^{\beta}R+c_2\,\Box R^{\alpha\beta}\bigg\}\;.
\end{align}
The model-dependence is captured by the coefficients. The ones  for  the $P$-odd operators also take into account a choice of scheme for the $\epsilon$-tensor in dimensional regularisation, which we do not need to specify (this is similar to Refs.~\cite{Elias:1982ea,Filoche:2022dxl} whereby one uses free parameters in order to remain independent of a specific $\gamma_5$-scheme).

\subsection{Diffeomorphism and Lorentz constraints}
There remains now to enforce the diffeomorphism and the Lorentz constraints \eqref{eq:diffeoLorentzconstraints} to 
deduce  relations on the coefficients in \eqref{eq:Talphabeta}.

Firstly, the reader may have noticed that we considered only symmetric operators in \eqref{eq:Talphabeta}. 
The reason is that since $\mathcal{T}^{\alpha\beta}_{ct}$ is symmetric, the Lorentz constraint is directly verified.
 
The diffeomorphism constraint $D_{\alpha}\mathcal{T}_{ct}^{\alpha\beta}=0$ leads to a system of equations. For the $P$-even part we obtain a non-trivial solution
\begin{alignat}{6}
\label{eq:7}
&4\,a_1+b_1&\;=\;&0 \;, \quad 
& & 4\,a_2+b_2+b_3&\;=\;&0\;, \quad 
& & 4\,a_3+b_4&\;=\;&0 \nonumber\;, \\
& 4\,a_1+a_2+a_4&\;=\;&0\;, \quad 
& & 8\,a_3-b_2 &\;=\;&0\;, \quad 
& & 2\,a_2+8\,a_3+c_2 &\;=\;&0\nonumber\;, \\  
&4\,a_4+c_1+c_2 &\;=\;& -12\,a_1-4\,a_2-4\,a_3  \;, \!\! \!\!\!\!\!\!\!\!\!\!\!\!\!\!   \!\! \!\!\!\!\!\!\!\!\!\!\!\!\!\! \!\! \!\!\!\!\!\!\!\!\!\!\!\!\!\!  & & && && &   
\end{alignat}
whereas for the $P$-odd part, there exists only the trivial solution
\begin{equation}
\label{eq:ee}
e=e_1=e_2=e_3=e_4=0\;.
\end{equation}
This is remarkable, since it means that the $\textit{finiteness}$ of the diffeomorphism and Lorentz anomalies implies the absence of parity violation in the EMT counterterms $\mathcal{T}_{ct}$.

\subsection{Consequences on the Weyl anomaly}

According to \eqref{eq:WeylTct}, the Weyl anomaly follows from taking the trace of $\mathcal{T}_{ct}$. Under the previous constraints, we obtain
\begin{align}
\label{eq:traceT+diffeo}
g^{(d)}_{\al\be}\mathcal{T}_{ct}^{\al\be} 
& = \frac{a_2+4\,a_3}{2}W^2 -\frac{a_2+2\,a_3}{2}E+\frac{3\,a_1+a_2+a_3}{3}R^2\;+ \\[0.1cm]
&\left( -(4\,a_1+a_2) -\frac{4}{d-4}(3\,a_1+a_2+a_3)\right) \Box R \;, \nonumber
\end{align}
where we expressed the result in terms of the Weyl tensor squared $W$ and the Euler density $E$ defined in Section~\ref{sec:intro}. The Weyl anomaly should be finite, hence this result is not yet complete.

\paragraph{Classically Weyl invariant theory}
In the case of a classically Weyl invariant theory, the anomaly must respect the Wess-Zumino consistency conditions\cite{Wess:1971yu}. We have to enforce them by hand, as they were not taken into account in our ansatz \eqref{eq:Talphabeta}. It is known that the Weyl tensor, the Euler density and $\Box R$ respect the Wess-Zumino consistency consitions, as opposed to the $R^2$-term\cite{Bonora:1985cq}, hence we must have
\begin{equation}
3\,a_1+a_2+a_3=0\;.
\end{equation}
This results in a finite Weyl anomaly constrained as
\begin{equation}
\label{eq:A2}
\mathcal{A}_{\mathrm{Weyl}}  = a_1\, R^2+a_2\,R_{\mu\nu}R^{\mu\nu}- (3\, a_1 +a_2) \,R_{\mu\nu\rho\sigma}R^{\mu\nu\rho\sigma}- (4\, a_1+a_2) \,\Box R\;,
\end{equation}
which depends on the two  parameters $a_1$ and $a_2$  which are model-dependent.
This is the result obtained by Duff\cite{DUFF1977334}, supplemented by 
the  additional constraint that  the $P$-odd $R\tilde R$-term  is absent. 

\paragraph{Classically Weyl non-invariant theory}
For an explicitly broken theory, the Weyl anomaly is not integrable but is given by the finite part of \eqref{eq:traceT+diffeo}, according to \eqref{eq:WeylTct}
\begin{equation}
\label{eq:A3}
\mathcal{A}_{\mathrm{Weyl}}  = a_1\, R^2+a_2\,R_{\mu\nu}R^{\mu\nu}+a_3\,R_{\mu\nu\rho\sigma}R^{\mu\nu\rho\sigma}- (4\,a_1+a_2) \,\Box R \;,
\end{equation}
with the absence of parity odd operators. We remark that the coefficient of the $\Box R$-term is fixed with respect to the others, even in an explicitly broken theory, which is a new observation. The $\Box R$ is considered ambiguous in the literature notably because of the possibility to include the counterterm
\begin{equation}
\label{eq:dS}
 S_{R^2} =\int\dd^4x\,\g\,\alpha\,R^2\;, \quad    \tensor{T}{^\mu_\mu} \supset  \frac{2}{\g}   g_{\mu\nu}\frac{\delta }{\delta g_{\mu\nu}}  S_{R^2} =-\frac{\alpha}{3}\Box R\; ,
\end{equation}
in $W_{ct}$ that alters the value of the $\Box R$ coefficient. However, since this counterterm is finite and independent from the quantum fields, it is just an explicit breaking, and does not enter in the anomaly $\mathcal{A}_{\mathrm{Weyl}}$ (see Ref~\cite{Prochazka:2017pfa} and Refs. in ~\cite{Larue:2023qxw} for more details).

\section{Conclusion}
\label{sec:conclusion}

Let us summarise our contribution to the 17th Marcel Grossmann Meeting.  
Motivated by a controversy in the literature we investigated the trace anomaly of a Weyl fermion. 
Using path integral methods (proper time regularisation of the determinant) we have shown that 
the Weyl anomaly of a free Weyl fermion is one-half of its Dirac counterpart  \eqref{eq:AWeyl} 
and does therefore not lead to anomalous parity violation~\cite{Larue:2023tmu}.
Moreover,  we have shown that any theory, whose symmetries is compatible with dimensional regularisation, 
does not allow for a $R\tilde R$- nor a $F \tilde F$-term in the trace anomaly~\cite{Larue:2023qxw}
(cf. \eqref{eq:A2} and \eqref{eq:A3} for the case without and with classical Weyl-breaking).\footnote{This excludes conclusions 
on supersymmetric theories as they are known to be at odds with dimensional regularisation.} 
The key-ingredient to obtain this model-independent result is a generic ansatz combined with 
the \emph{finiteness} of the gauge, the  Lorentz and the diffeomorphism anomalies. 
In addition it was shown that the $\Box R$-term is unambiguous which is often stated to the contrary in the literature.

\section{Acknowledgments}
The work of RL and JQ is supported by the project AFFIRM of the Programme National GRAM of CNRS/INSU with INP and IN2P3 co-funded by CNES and by the project EFFORT supported by the programme IRGA from UGA. RL, JQ and RZ acknowledge the  support of CERN associateships. The work of RZ is supported by the  STFC Consolidated Grant, ST/P0000630/1. 
Many manipulations were carried out with the help of Mathematica and the package xAct \cite{xAct}.

\appendix{Diffeomorphism and Lorentz transformations}
\label{app:transfo}

We denote by $\delta^d_\xi$ the infinitesimal diffeomorphism (active coordinate transformation) of parameter $\xi_\mu(x)$.
The transformation of the weight $-1/2$ fermions\cite{Toms:1986sh}, that enter in the diffeomorphism-invariant path integral measure \eqref{eq:WSL}, and of the vierbein, vierbein determinant, spin-connection and gauge field are given by
\begin{alignat}{4}
&\delta^{d}_\xi\tilde\psi&\;=\;&\xi^\mu\pa_\mu\tilde\psi+\frac{1}{2}(\pa_\mu\xi^\mu)\tilde\psi
\;, \quad   & &  \delta^{d}_\xi\tilde{\bar\psi}&\;=\;&\xi^\mu\pa_\mu\tilde{\bar\psi}+\frac{1}{2}(\pa_\mu\xi^\mu)\tilde{\bar\psi} \;, \nonumber \\[0.1cm]
&\delta^{d}_\xi\tensor{e}{^\mu_a}&\;=\;&\xi^\nu\pa_\nu\tensor{e}{^\mu_a}-\tensor{e}{^\nu_a}\pa_\nu\xi^\mu\;, \quad   & &
\delta^{d}_\xi e&\;=\;&(\pa_\mu e\,\xi^\mu) = e (D_\mu\xi^\mu)  \;, \nonumber \\[0.1cm]
&\delta^{d}_\xi\omega_\mu&\;=\;&\xi^\nu\pa_\nu\omega_\mu+\omega_\nu\pa_\mu\xi^\nu \;, \quad   & &
\delta^d_\xi A_\mu&\;=\;&\xi^\nu \pa_\nu A_\mu+A_\nu \pa_\mu \xi^\nu \;.
\end{alignat}
Similarly, under an infinitesimal Lorentz transformation $\delta^L_\alpha$ of parameter $\alpha_{ab}(x)=-\alpha_{ba}(x)$, we have
\begin{alignat}{4}
&\delta^{L}_\alpha\tilde\psi &\;=\;& -\frac{1}{2}\alpha_{ab}\Sigma^{ab}\, \tilde\psi 
\;,\quad \delta^{L}_\alpha\tilde{\bar\psi}=\frac{1}{2}\alpha_{ab}\,\tilde{\bar\psi} \Sigma^{ab}
\;, \quad \delta^{L}_\alpha\tensor{e}{^\mu_a}=\tensor{e}{^\mu_b}\tensor{\alpha}{^b_a}
\;, \quad   & & \delta^{L}_\alpha e&\;=\;& 0 \;, \nonumber \\[0.1cm]
&\delta^{L}_\alpha \omega_\mu &\;=\;& \frac{1}{2}[D_\mu,\alpha_{ab}\Sigma^{ab}]=\frac{1}{2}\Sigma^{ab}[D_\mu,\alpha_{ab}] \;, \quad     && \delta^{L}_\alpha A_\mu&\;=\;0&  \;,
\end{alignat}
where $\Sigma^{ab}=\frac{1}{4}[\gamma^a,\gamma^b]$.

For completion, we also give the transformation of the Fadeev-Popov ghosts and gauge fixing terms for the action
\begin{equation}
S=\int\dd^4x\g\left( -\frac{1}{4}g^{\mu\rho}g^{\nu\sigma}F_{\mu\nu}F_{\rho\sigma}-(\partial_\mu B)g^{\mu\nu}A_\nu+i(\partial_\mu\bar c)g^{\mu\nu}\partial_\nu c\right)\;,
\end{equation}
which are
\begin{equation}
\delta^W_\sigma c=0\;\quad\quad\delta^W_\sigma \bar c= -2\sigma\bar c\;\quad\quad \delta^W_\sigma B= -2\sigma B\;.
\end{equation}
The diffeomorphism and Lorentz transformations are those of scalar fields
\begin{equation}
\delta^d_\xi \phi=\xi^\mu\partial_\mu\phi\;\quad\quad\delta^L_\alpha\phi=0\;.
\end{equation}

\bibliographystyle{ws-procs961x669}
\bibliography{biblio}
\end{document}